\documentclass[10pt,conference,letterpaper]{IEEEtran}
\IEEEoverridecommandlockouts
\usepackage{cite}
\usepackage{amsmath,amssymb,amsfonts}
\usepackage{algorithmic}
\usepackage{graphicx}
\usepackage{textcomp}
\usepackage{xcolor}
\usepackage{subfig}
\usepackage[letterpaper, right=0.68in,left=0.68in,top=0.75in,bottom=1.027in]{geometry}
\usepackage{tikz}
\usepackage{textcomp}
\usepackage{hyperref}

\newcommand\copyrighttext{%
 \textcopyright 2024 IEEE. Personal use of this material is permitted.
	Permission from IEEE must be obtained for all other uses, in any current or future
	media, including reprinting/republishing this material for advertising or promotional
	purposes, creating new collective works, for resale or redistribution to servers or
	lists, or reuse of any copyrighted component of this work in other works.
	}
\newcommand\copyrightnotice{%
	\begin{tikzpicture}[remember picture,overlay]
	\node[anchor=south,yshift=10pt] at (current page.south) {\fbox{\parbox{\dimexpr\textwidth-\fboxsep-\fboxrule\relax}{\copyrighttext}}};
	\end{tikzpicture}%
}

\def\BibTeX{{\rm B\kern-.05em{\sc i\kern-.025em b}\kern-.08em
    T\kern-.1667em\lower.7ex\hbox{E}\kern-.125emX}}
\begin{document}

\title{A Signature Based Approach Towards Global Channel Charting with Ultra Low Complexity%: Towards Unsupervised Learning based Localizations\\
}
\author{\IEEEauthorblockN{Longhai Zhao,  Yunchuan Yang, Qi Xiong, He Wang, Bin Yu, Feifei Sun,  and Chengjun Sun}
	\IEEEauthorblockA{\textit{Advanced Research \&\ Standard Team,} 
		\textit{Samsung Research Institute China – Beijing (SRC-B)},
		Beijing, P. R. China \\
		%	\textit{$^\dagger$R \& D Team, Networks Business Division, Samsung Electronics}, Suwon, South Korea\\
		\{longhai.zhao, yc0301.yang, q1005.xiong, h0809.wang, bin82.yu, feifei.sun, chengjun.sun\}@samsung.com}
}

%\and
%\and
%\and
%\and
%\and
\maketitle
\copyrightnotice
\begin{abstract}
Channel charting, an unsupervised learning method that learns a low-dimensional representation from channel information to preserve geometrical property of physical space of user equipments (UEs), has drawn many attentions from both academic and industrial communities, because it can facilitate many downstream tasks, such as indoor localization, UE handover, beam management, and so on. However, many previous works mainly focus on charting that only preserves local geometry and use raw channel information to learn the chart, which do not consider the global geometry and are often computationally intensive and very time-consuming. Therefore, in this paper, a novel signature based approach for global channel charting with ultra low complexity is proposed. By using an iterated-integral based method called signature transform, a compact feature map and a novel distance metric are proposed, which enable 
%The proposed approach utilizes signature transform, an iterated-integral based transform, to extract a useful and compact feature map and to design a novel distance metric, resulting in a 
 channel charting with ultra low complexity and preserving both local and global geometry. We demonstrate the efficacy of our method using synthetic and open-source real-field datasets.
\end{abstract}


\section{Introduction}
%This document is a model and instructions for \LaTeX.
%Please observe the conference page limits. 
%
%1. The problem is important.
%
%
%2. The problem is hard.
%
%
%3. The problem is sovled by the proposed method.

%
%Channel charting is to learn faithful low-dimensional representations of high dimensional wireless channels from user equipments (UE) with different locations to a set of base stations.  By faithful representations it means that close UEs in spatial locations have similar low-dimensional representations, that is the neighborhood relationships between UEs are kept and encoded in the low-dimensional representations.
%Accurate location information is in great demand for numerous vertical services. However, 
Channel charting (CC) is an unsupervised learning method that uses channel information (CI) to extract a low-dimensional embedding (i.e., the chart) that preserves geometrical structure of physical space of user equipments (UEs), which facilitates a broad distance-based or location-based applications, such as handover \cite{cc}, indoor localization \cite{frauhofer},  beam management \cite{bt}, pilot management \cite{pilot}, and so on. Thus, CC is an attractive and promising approach to enable many applications in beyond 5G and 6G systems. Roughly speaking, CC most often  consists of three main steps: 1.) construct a suitable feature map of the CI, 2.) design a distance metric for the CI or feature map in such a way that the distance between two feature maps is proportional to the physical distance between the corresponding UEs, 3.) use a dimensionality reduction method to obtain the chart.\footnote{For some non-parametric approaches, such as principal component analysis, no need to specify a distance metric explicitly.}% first two steps play a crucial role for a successful charting. 

%

%Accurate positioning in scenarios with predominant multi-path propagation and non-line-of-sight (NLOS) is a challenging task, in which conventional positioning methods like trilateration based on time of arrival (TOA) usually perform poorly in these scenarios. In recent years, artificial intelligence or machine learning (AI/ML) based approaches have shown promising results on NLOS-heavy scenarios \cite{b1}\cite{b2}. On the other hand, AI/ML based positioning has been approved in the 3rd Generation Partnership Project  (3GPP) as one of the three representative cases to apply AI in wireless networks\cite{b3}. After almost a year's study, the 3GPP concludes the observations suggesting that AI/ML can magnificently improve positioning for both direct AI positioning and AI assisted positioning\cite{bb4}.

However, in CC studies so far, the raw channel information, e.g., channel impulse response (CIR) or channel frequency response (CFR), and their simple variants are still commonly used as feature maps to learn the chart. For instance, channel features in angular domain are proposed for CC in \cite{cc}\cite{btri}, which do not apply for scenarios with single-antenna base stations (BSs). The magnitude of CIR is used as feature map in \cite{frauhofer} and the  magnitude of truncated CIR is used in \cite{taner}. All existing feature maps used for CC are still very high-dimensional owing to the increasing sampling rate in the spatial, frequency, and time domains, which increase the complexity of subsequent procedures for CC, e.g., the pairwise distance calculation, the operations for training and inference of neural networks. Since CC is a similarity/dissimilarity learning based method and the number of pairwise distance is proportional to the squared number of training samples, more training time is needed for neural network based CC. Therefore, a compact and suitable feature map is very appealing for practical implementations of CC. But there is still a  lack of  systematic ways for deriving a feature map that is low-dimensional, sampling-rate-agnostic and suitable for CC. 

%\vspace{-0.04cm}
When using parametric approaches, such as Siamese network or triplet network, to learn the chart, designing a faithful distance metric is crucial for a successful CC. However, the UE trajectory or distribution in real environments often exhibits a geometrical structure that is locally linear and globally nonlinear, which poses challenges to design a simple metric that is faithful for both local and global geometry. Using the norm of channel features in angular domain \cite{cc}, time domain \cite{frauhofer},  or timestamps of consecutive CI \cite{btri} as a metric only preserves the local geometry, which makes the resulting CC inappropriate for tasks like indoor localization. In order to obtain a distance metric that is suitable for both local and global geometry, the CIR geodesic distance is proposed in \cite{frauhofer}. However, to generate the pairwise CIR geodesic distance matrix, one needs to specify a suitable neighborhood number, which is a non-trivial task and dependent on particular UE trajectories or distributions. In addition, constructing the neighborhood graph and calculating the geodesic distance are also computationally intensive. 
\addtolength{\topmargin}{-0.04in}
\vspace{-0.06cm}

Therefore, to address these issues in the first two steps of CC, a signature based approach that generates a global chart with ultra low complexity is proposed in this paper. Instead of using truncation \cite{taner} or downsampling \cite{btri} to reduce the number of channel features, we take a totally opposite way, in which the dimensionality of raw discrete CIRs is firstly lifted to form  continuous functions and then reduced significantly to obtain a compact feature map by using signature transform, an iterated-integrals based transformation.  The signature transform is a nonlinear mapping from a path (continuous mapping) to an infinite sequence called signature, which was originally introduced by \cite{Chen} and has been used in finance\cite{b17},  rough path theory \cite{b16}, machine learning\cite{sigkernel}, and so on.
%\vspace{0.02in} 
The signature provides a well summary of the path and has meaningful interpretations from the geometrical or statistical viewpoint. With the proposed feature map, a signature based principal component analysis (SPCA) is proposed for CC, in which the dimensionality of the covariance matrix is very small and only proportional to the number of BSs. For parametric approaches, we propose a signature based Siamese network (SSN) for CC. In order to train the network efficiently in terms of both performance and complexity, a novel distance metric that is valid for both local and global geometry is also proposed, without calculating geodesic distance or constructing neighborhood graph. We evaluate the proposed methods on a synthetic dataset and two open-source real-field datasets.

\begin{figure*}[ht!]
	\centering
	\includegraphics[width=16cm]{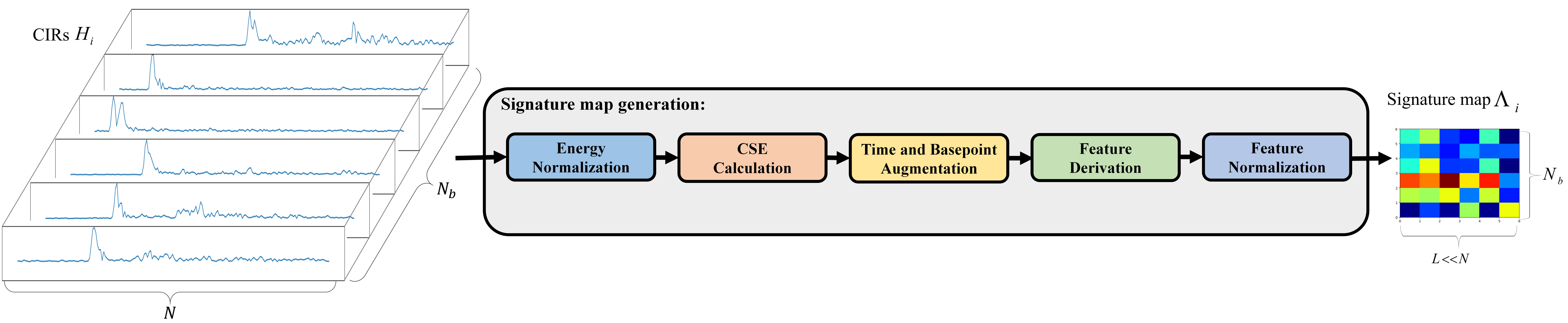}
	\caption{An illustration of the signature map generation.}
	\label{figcca}
	\vspace{-0.5cm}
\end{figure*}

The outline of the paper is as follows. The preliminaries of path signature are introduced in Section II. Then, the proposed signature-based approach is presented in Section III with the introduction of feature map derivation, SPCA and SSN based CC. Eventually, the experimental results are given and demonstrate the superiority of the proposed approach.
\section{Preliminaries of path signature}
A $d$-dimensional path (parameterized by $\tau$) is a continuous mapping from a real interval $[a,b]$ to $\mathbb{R}^d$, and can be denoted as $X(\tau)=\{X_1(\tau),....,X_d(\tau)\}:[a,b]\rightarrow\mathbb{R}^d$. The signature of $X(\tau)$ is defined as the collection of iterated integrals, i.e.,
\begin{equation}\label{eq:sig_def}
\mathbf{S}(X)=(1,S(X)_1,\cdots,S(X)_d,\cdots,S(X)_{i_1,i_2,\cdots,i_k},\cdots)
\end{equation}
where "$1$" as the first term of $\mathbf{S}(X)$ denotes the $0$-th level signature, and $S(X)_{i_1,i_2,\cdots,i_k}$ denotes one of the $k$-th level ($k$ is an integer and $k\geq 1$) signature features and given by \cite{sigkernel}
\begin{equation}\label{eq:sdef}
	S(X)_{i_1,i_2,\cdots,i_k}=\mathop{\int\cdots\int}\limits_{a<\tau_1<\cdots<\tau_k<b}\text{d}X_{i_1}(\tau_1)\cdots\text{d}X_{i_k}(\tau_k)
\end{equation}
%
%\begin{equation}
%S(X)_{i_1,i_2,\cdots,i_k} = \mathop{\int\cdots\int}\limits_{a<\tau_1<\cdots<\tau_k<b}\prod_{m=1}^k\frac{\text{d}X_{i_m}}{\text{d}t}(\tau_m)\text{d}\tau_1\text{d}\tau_2\cdots \text{d}\tau_k
%\end{equation}
where $i_k\in\{1,2,\cdots,d\}$, and the integral can be understood as Riemann integral with $\text{d}X_{i_k}(\tau_k)=\frac{\text{d}X_{i_k}}{\text{d}\tau}(\tau_k)\text{d}\tau_k$ when the underlying path is differentiable.
%\addtolength{\topmargin}{0.02in}

Due to the nature of integral, each signature feature can be interpreted as a geometrical item like length, area, volume, and so on. Moreover, each signature feature can also be viewed as a statistical feature. For example, let $\tau$ be a random variable following uniform distribution over $[a,b]$, then the first and second level signature features are given by
\begin{equation}
	S(X)_i=\int_a^{b}\text{d}X_i(\tau)\propto \mathbb{E}_\tau[\dot{X}_i(\tau)]
\end{equation}
\begin{equation}
\small
	S(X)_{i,j}=\mathop{\int\int}\limits_{a<\tau_1<\tau_2<b}\text{d}X_i(\tau_1)\text{d}X_j(\tau_2)\propto\mathbb{E}_{\tau_1<\tau_2}[\dot{X}_i(\tau_1)\dot{X}_j(\tau_2)]
\end{equation}
where $\dot{X}_i(\tau)=\frac{\text{d}X_i(\tau)}{\text{d}\tau}$, $i$,  $j\in\{1,2,\cdots,d\}$. Thus, $S(X)_i$ measures the averaged change of random variable $X_i$ over the interval $[a,b]$.  $S(X)_{i,j}$ measures the covariance of changes between $X_i$ and $X_j$ at two ordered points, which makes it different from ordinary statistics like moments that do not have any ordering constraints. 
%\vspace{0.02in}
\addtolength{\topmargin}{-0.03in}

The signature has some nice properties that make it very suitable for CC. For example, the mapping from the path to its signature is continuous, injective under a mild condition, and the signature can approximate the path arbitrarily well by linear combination \cite{sigkernel}. Most importantly, the ability to preserve the ordering information for sequential data makes the signature very suitable for distance-based learning algorithms.

It is also worth to note that the features of signature are not independent from each other. To eliminate the redundant information of the signature, a more compact form can be obtained by its logarithm, which is defined as \cite{Chen}
\begin{equation}\label{eq:logsdef}
\log(\mathbf{S})\triangleq \sum_{m=1}^\infty\frac{(-1)^{(m-1)}}{m}(\mathbf{S}-\mathbf{1})^{\otimes m}
\end{equation}
where $\mathbf{1}=(1,0,0,\cdots)$, and $\otimes$ denotes the tensor product over non-commutative tensor algebra\cite{b16}. For the remainder of this paper, we will refer to the feature obtained by \eqref{eq:logsdef} as the signature. Moreover, a truncation of the signature is usually applied for practical usage.

\section{The proposed signature based CC}
Consider a single-input single-output (SISO) communication system in which UEs communicate to $N_b$ BSs.\footnote{The proposed method, however, can be applied to multi-antenna scenarios, in which $N_b$ denotes the number of antennas from all BSs. Alternatively, one can construct a path in the angular domain and generate a corresponding signature map, in particular when the spatial dimensionality is predominant.} We have a dataset with $D$ samples of raw CIRs for all links between BSs and any UE, which can be denoted as a set of matrices $\{\boldsymbol{H}_i\}$ with $\boldsymbol{H}_i$ being a $N_b\times 2N$ real matrix, $i \in\mathcal{D}:=\{1,2,...,D\}$ is the sample index and $N$ is the number of taps.\footnote{We use the tap to denote the more commonly known terminology of  multipath-component. Although that each CIR is assumed to have the same length, the proposed approach can be applied directly to a dataset that has variable-length CIRs.} With this dataset, the objective of CC is to learn a low-dimensional embedding, i.e., the channel chart, such that the spatial geometry of actual UE locations is preserved in the chart.
%\footnote{We assume that each CIR has the same length, but the proposed approach can also be applied directly to a dataset that has variable-length CIRs.}
%
%An illustration of the proposed method is shown in Fig.\ref{fig_cc}, which includes the signature map generation, SPCA based and SSN based CC. 
To achieve this goal, we first transform each $\boldsymbol{H}_i$ into a $N_b\times L$ matrix $\boldsymbol{\Lambda}_i$, called signature map, with $L<<N$, as shown in Fig.\ref{figcca}. Then we use this signature map to perform a further dimensionality reduction (DR) to obtain the final chart $\boldsymbol{o}_i\in\mathbb{R}^2$ or $\mathbb{R}^3$.  Although this further DR can be accomplished by many approaches, 
 such as Isomap \cite{Isomap}, t-distributed stochastic neighbor embedding (t-SNE) or Laplacian eigenmap, these approaches cannot be easily used to predict on unseen data. Therefore, the SPCA and SSN are proposed for the DR from the signature map to the chart. The details of the proposed method are given below.%The details of the signature map generation and SPCA or SSN based CC are given as follows.

% the proposed signature map can also be used for CC in combination with other approaches,
%
% Specifically,
%
%For non-parametric approaches, the vectorized signature map can be used as the input of PCA to obtain a channel chart, namely the SPCA. Note that one only needs to calculate the singular value decomposition (SVD) of a $N_bL\times N_bL$ matrix, which is significantly faster than PCA with raw CIRs. As will be shown in Section \ref{sec4}, a very small length $L\leq6$ of signature is enough to generate a good chart. Moreover, % For parametric approaches, we propose a SSN based CC, as shown in Fig.\ref{fig_cc_c}. With the reduced input feature number, the computational complexity and parameter number of the neural network can be reduced significantly. Moreover, a novel distance metric is proposed to calculate the pairwise distance matrix, which doesn't need to construct neighborhood graph or calculate geodesic distance and still show good results on the learned CC. The details of the proposed method is given in the following subsections.

%With the more compact feature map,  a shallow Siamese network is 

\subsection{Signature map generation}
To obtain the signature map, one needs firstly to construct a path from the raw discrete sequence. A path, constructed simply by linear interpolating the CIR sequence, usually doesn't suit the purpose of CC well. Therefore, we propose the cumulative sum of energy (CSE) path with time and basepoint augmentations. The detailed steps are given below.
%, and then use the resulting path to get the signature map
 % Therefore, some preprocessing are needed 

%Each raw of $\boldsymbol{H}_i$ is a discrete-time baseband impulse response of the corresponding channel. The first step is to normalize the total energy of the CIR, which can reduce the scaling problem of large-scale fading in the CIR. The energy normalized CIR can be written as an ordered sequence 
%\begin{equation}
%	\{\boldsymbol{h}_n\}_{n=1}^{N}=((a_1,b_1),(a_2,b_2),\cdots,(a_N,b_N))
%\end{equation}
%%\begin{equation}
%%h(t)=\sum_{n=1}^N(a_n+jb_n)\delta(t-t_n)%\quad\quad\quad n\in {1,2,...,N}
%%\end{equation}
%where $\boldsymbol{h}_n$ is a 2-tuple vector with $a_n\in\mathbb{R}$ and $b_n\in\mathbb{R}$ representing the real and imaginary parts of the $n$-th tap, respectively. % $\delta(\cdot)$ is the Dirac delta function.

%
%By normalized CSE path, we mean two things, the first to normalized the CIR total energy, and the second is to normalize each signature features. The feature derivation consists of the following main steps:
\begin{itemize}
	\item \textbf{Energy normalization}. The first step is to normalize the total energy of CIR, which can solve the scaling problem owing to the large-scale fading. The resulting CIR can be written as an ordered sequence,
	\begin{equation}
		\{\boldsymbol{h}_n\}_{n=1}^{N}=((a_1,b_1),(a_2,b_2),\cdots,(a_N,b_N))
	\end{equation}
	%\begin{equation}
	%h(t)=\sum_{n=1}^N(a_n+jb_n)\delta(t-t_n)%\quad\quad\quad n\in {1,2,...,N}
	%\end{equation}
	where $\boldsymbol{h}_n$ is a 2-tuple vector with $a_n, b_n\in\mathbb{R}$ representing the real and imaginary parts of the $n$-th tap, satisfying $\sum_{n=1}^Na_n^2+b_n^2=1$.
	\item  \textbf{Calculation of CSE sequence}. The second step is to obtain the CSE sequence, which is given by 
	\begin{equation}
	\{c_n\}_{n=1}^N= (c_1,c_2,\cdots,c_N)
	\end{equation}
	where $c_n =\sum_{k=1}^np_k$ and $p_k=a_k^2+b_k^2$. Note that $c_N=1$ after energy normalization of CIR. The CSE sequence is non-decreasing, which makes the mapping from the CSE path to signature become injective and the resulting features of signature have meaningful interpretations from geometrical or physical viewpoint.
	% which can be used to form a non-decreasing continuous function with its signature having nice interpretation from geometrical or statistical viewpoint.
	% Moreover, the phase 
%	\begin{equation}
%\quad\quad n\in\{1,2,\cdots,N\}
%	\end{equation}
	\item \textbf{Time augmentation}. The time augmentation intends to lift the sequence dimensionality and emphasize the time information, since the time information is crucial for CC in case of SISO scenarios. The operation for time augmentation is simply adding the timing information for each tap, i.e., 
	\begin{equation}
	\{\boldsymbol{c}_n^{(aug)}\}_{n=1}^N=((c_1,t_1),(c_2,t_2),\cdots,(c_N,t_N))
	\end{equation}
	where $t_n$ is the time for the $n$-th tap. For CIRs measured with uniform sampling, $\{t_n\}_{n=1}^N$ is an evenly spaced sequence and depends on the sampling rate. If all samples in the dataset use the same sampling rate and have the same number of CIR lengths, $\{t_n\}_{n=1}^N$ can be set as $\{n/N\}_{n=1}^N$, in virtue of the signature property of  invariance to reparameterization.
	\item \textbf{Basepoint augmentation}.  The operation for basepoint augmentation is simply adding a 2-tuple zero vector at the start, i.e.,
	\begin{equation}
	\{\boldsymbol{c}_n^{(base)}\}_{n=0}^{N}=((c_0,t_0),(c_1,t_1),(c_2,t_2),\cdots,(c_N,t_N))
	\end{equation}
	Here $c_0\triangleq 0$ and $t_0\triangleq 0$.
	The purpose of basepoint augmentation is to make an alignment for the starting point of each time-augmented CSE sequence.  By aligning the basepoint, the integrity of time and energy information is preserved, especially for CIRs with the first tap having the highest energy. Because some signature features depend on the difference between the endpoint and starting point of the CSE sequence, the possibly important information of the first tap of CIRs will be lost if there is no basepoint augmentation.
	\item \textbf{Feature derivation.}
	With the time and basepoint augmented CSE sequence, a two-dimensional CSE path $X:[t_0,t_N]\rightarrow \mathbb{R}^2$ can be constructed by linear interpolation,  which is given by 
	\begin{equation}
	X(t)=\left\{\begin{array}{c}
	X_1(t)=c(t)\\
	X_2(t)=t
	\end{array}\right.
	\end{equation}
	where $c(t)=\frac{c_n-c_{n-1}}{t_n-t_{n-1}}(t-t_{n-1})+c_{n-1}$ (for $t_{n-1}\leq t\leq t_n$)  is a piece-wise linear function of time. 
%	\begin{equation}
%		c(t) =\frac{c_n-c_{n-1}}{t_n-t_{n-1}}(t-t_{n-1})+c_{n-1},\quad t_{n-1}\leq t\leq t_n
%	\end{equation}
	 Then its truncated signature can be obtained by \eqref{eq:sdef} and \eqref{eq:logsdef}. A practical implementation for calculating the signature can be found in \cite{iisig}. We denote the obtained signature as a vector $\boldsymbol{\lambda}\in\mathbb{R}^L$, where $L$ denotes the number of features and depends on the level at which the truncation is performed. Note that the complexity of calculating the signature is linear with respect to $N$, which usually does not bring much computational cost,  given the significant complexity reduction on the subsequent procedures.
	
	%with 8 entries.
	% We further remove the two features of the first level which are uninformative, resulting a vector $\boldsymbol{\lambda}\in\mathbb{R}^6$. 

%		\begin{flalign}
%	& \lambda_{1,2} = \frac{1}{2}\sum_{n=1}^Np_n(t_N-t_n-t_{n-1})& \label{eq:lambda12}
%	\end{flalign}
%	\vspace{-0.4cm}
%	\begin{align}
%	\lambda_{1,1,2}&=\sum_{n=1}^N(t_n-t_{n-1})(\frac{1}{2}c_{n-1}^2+\frac{1}{2}c_{n-1}(p_n-c_N)+\frac{1}{6}p_n^2\quad\notag \\
%	&-\frac{1}{4}c_Np_n)+\frac{1}{12}c_N^2t_N\label{eq:lambda112}
%	\end{align}
%	\vspace{-0.5cm}
%	\begin{align}\label{eq:lambda122}
%	\lambda_{1,2,2}&= \frac{1}{12}\sum_{n=1}^Np_n(t_N^2+2t_n^2+2t_{n-1}^2+2t_nt_{n-1}\quad\quad\quad\quad\quad\notag\\
%	&-3t_N(t_n+t_{n-1}))
%	\end{align}
%	\vspace{-0.5cm}
%	\begin{align}
%	\lambda_{1,1,1,2}&=\sum_{n=1}^N(t_n-t_{n-1})(\frac{1}{6}c_{n-1}^3+\frac{1}{4}c_{n-1}^2(p_n-c_N)\notag\\\label{eq:lambda1112}
%	&+\frac{1}{12}c_{n-1}(c_N-p_n)(c_N-2p_n)+\frac{1}{24}p_n(p_n-c_N)^2)
%	\end{align}
%	\vspace{-0.5cm}
%	\begin{align}
%	\lambda_{1,1,2,2}& = \frac{1}{24}c_N^2t_N^2+\frac{1}{24}\sum_{n=1}^Np_n(t_n^2(2c_N-p_n)+t_{n-1}^2(2c_N\notag\\
%	&-3p_n)+2t_nt_{n-1}(c_N-p_n)-2t_Nt_n(2c_N-p_n)\notag\\
%	&-4t_{n-1}t_N(c_N-p_n))+\sum_{n=1}^Np_n\times\notag\\\label{eq:lambda1122}
%	&\sum_{k=1}^{n-1}(\frac{1}{4}p_kt_N(t_k+t_{k-1})-\frac{1}{6}p_k(t_k^2+t_kt_{k-1}+t_{k-1}^2))
%	\end{align}
%	\begin{align}
%	\lambda_{1,2,2,2}&=\frac{1}{24}\sum_{n=1}^Np_n(2t_Nt_{n-1}^2-(t_n+t_{n-1})((t_N-t_n)^2\notag\quad\\
%	&+t_{n-1}^2))\label{eq:lambda1222}
%	\end{align}
	\item \textbf{Feature normalization}. The obtained signature is further normalized across the dataset so that each feature is zero-mean and unit-variance.\footnote{Note that, in addition to improving the training efficiency, this step can also reduce the homoskedastic-noise effect from noisy CI. In case of datasets  with heteroskedastic noise, path averaging or signature averaging across consecutive CI can be used for denoising, and other advanced methods, such as \cite{kernelPCA}, work as well.} %e, which can improve the performance for PCA and the training efficiency for SSN based CC.
%	 The $k$th entry of the normalized signature $\hat{\boldsymbol{\lambda}}$ is given by 	$\hat{\lambda}_k=\frac{\lambda_k-\mu_k}{\sigma_k}$.
%%	\begin{equation}
%%		\hat{\lambda}_k=\frac{\lambda_k-\mu_k}{\sigma_k}
%%	\end{equation}
%	where $\mu_{k}$ and $\sigma_{k}$ is the mean and standard deviation for corresponding feature, respectively, and $k\in\{1,2,...,L\}$.% $i\in\{1,2,..,D\}$, and $m\in\{1,2,..,N_b\}$.
%	\item  \textbf{Feature derivation}
\end{itemize}
%\addtolength{\topmargin}{-0.04in}

By performing the above operations for each CIR from each BS,  the raw channel information of each sample $\boldsymbol{H}_i$ is transformed into a more compact signature map $\boldsymbol{\Lambda}_i$, which is used as the feature map for the subsequent CC.

\subsection{SPCA and SSN}
 After obtaining the signature map, the inner product of the signature becomes a faithful similarity measure, and CC by PCA becomes attractive because of its practical implementation, computational efficiency, and the ability to inference on unseen data. The SPCA can also be viewed as a kernelized PCA.  Moreover, we can further select a subset of vectors from the signature map, by which the computational complexity can be further reduced without degrading the CC performance seriously, as will be shown in Section \ref{sec4}.
 
\begin{figure}[t]
	\vspace{0.05in}
	\centering
	\includegraphics[width=8cm]{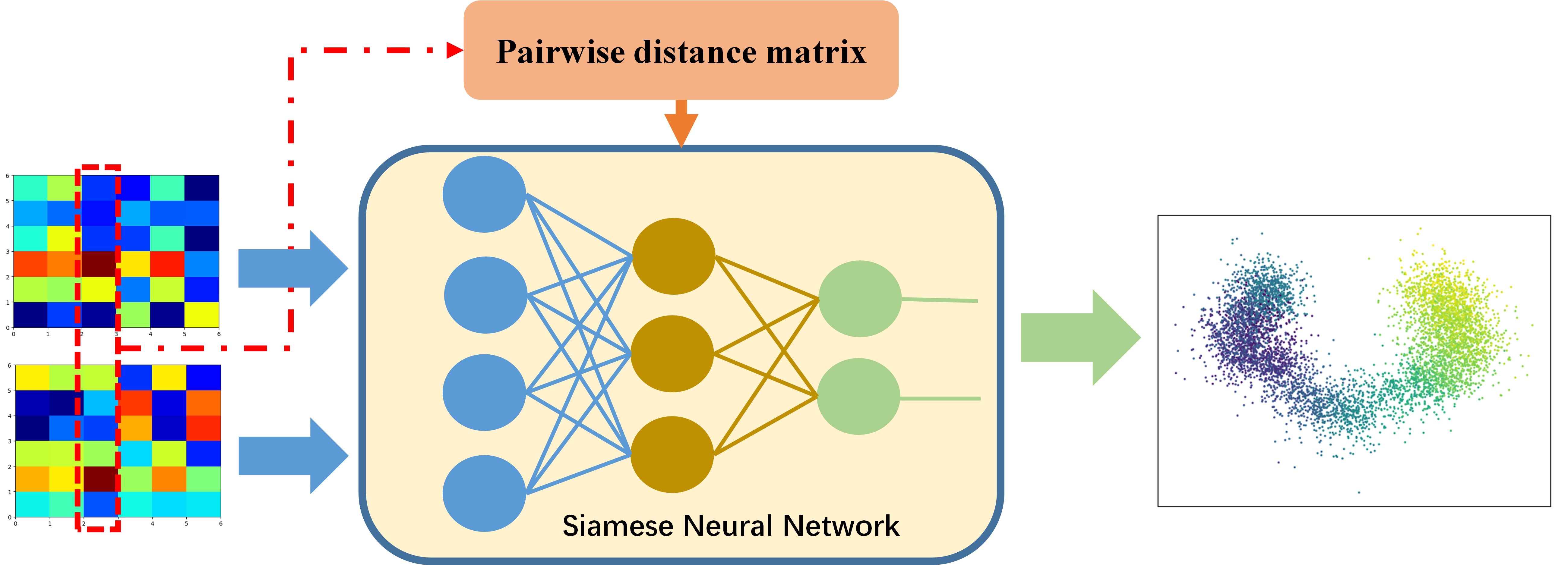}%\label{fig_cc_a}
	\caption{An illustration of SSN based CC.}
	\label{fig_cc_sia}
	\vspace{-0.5cm}
\end{figure}

With the obtained signature map,  Siamese network can be used to learn the chart, as shown in Fig.\ref{fig_cc_sia}. To train the network, We propose the following distance metric for any pair of signature maps,
\vspace{-0.03in}
\begin{equation}\label{eq:dist_def}
	\rho(\boldsymbol{\Lambda}_i,\boldsymbol{\Lambda}_j)\triangleq \rVert\boldsymbol{s}_i-\boldsymbol{s}_j\rVert_1
\end{equation}
where  $\boldsymbol{s}_i=[s^{(i)}_1,s^{(i)}_2,\cdots,s^{(i)}_m,\cdots,s^{(i)}_{N_b}]^T\in\mathbb{R}^{N_b}$, $\rVert\cdot\rVert_1$ denotes the $l^1$ norm, and $s^{(i)}_m$ denotes one of the signature feature from the $i$-th sample and $m$-th BS, which is given by
%\begin{align}\label{eq:lambda122}
%s_m^{(i)}&= \frac{1}{12}\sum_{n=1}^Np_n(t_N^2+2t_n^2+2t_{n-1}^2+2t_nt_{n-1}\quad\quad\quad\quad\quad\notag\\
%&-3t_N(t_n+t_{n-1}))
%\end{align}
\begin{equation}\label{eq:lambda122}
\small{
s_m^{(i)}= \frac{1}{12}\sum_{n=1}^Np_n(t_N^2+2t_n^2+2t_{n-1}^2+2t_nt_{n-1}-3t_N(t_n+t_{n-1}))
}
\end{equation}
%\begin{equation}
%
%\end{equation}
%We define the pairwise distance matrix 	$\boldsymbol{\Sigma}\in\mathbb{R}^{D\times D}$ of signature map as the $l_1$ norm of the vector $\boldsymbol{s}_i$, i.e.
%\begin{equation}
%[\boldsymbol{\Sigma}]_{i,j}=d_s(\boldsymbol{\Lambda}_i,\boldsymbol{\Lambda}_j) = 
%\end{equation}
%\begin{figure}[htbp]
% 	\centerline{\includegraphics[width=0.8\linewidth]{heatmap_dist_mat.png}}
% 	\caption{Heatmaps for the pairwise distance matrices, generated from true radio enviroments, the proposed signature distance, and the CIR geodesic distance.}
% 	\label{figheat}
%\end{figure}    
%\vspace{0.05in}
 \begin{figure}[ht!]
	\vspace{-0.8cm}
	\centering
	\subfloat[Distance of true locations]{
		\includegraphics[width=4cm,height=2.5cm]{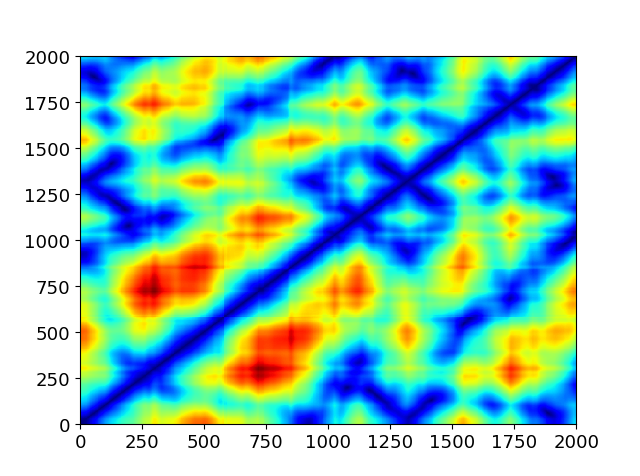}}
	\subfloat[The proposed signature distance]{
		\includegraphics[width=4cm,height=2.5cm]{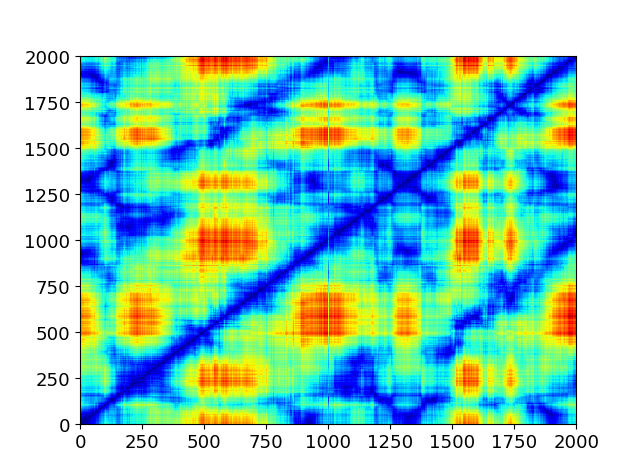}}
	
	\vspace{-0.3cm}
	\subfloat[CIR geo. distance w/ $K=5$]{
		\includegraphics[width=4cm,height=2.5cm]{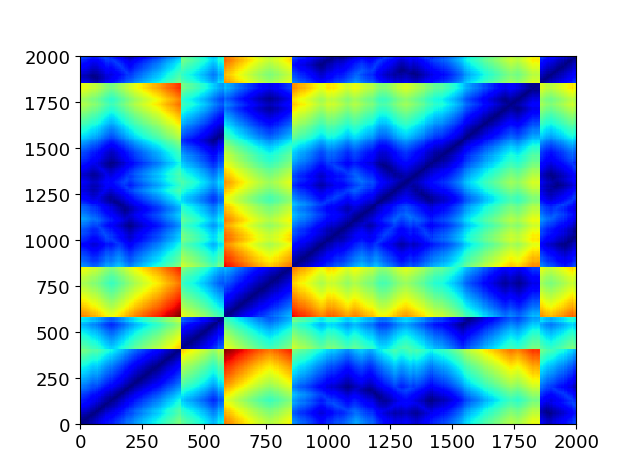}}
	\subfloat[CIR geo. distance w/ $K=15$]{
		\includegraphics[width=4cm,height=2.5cm]{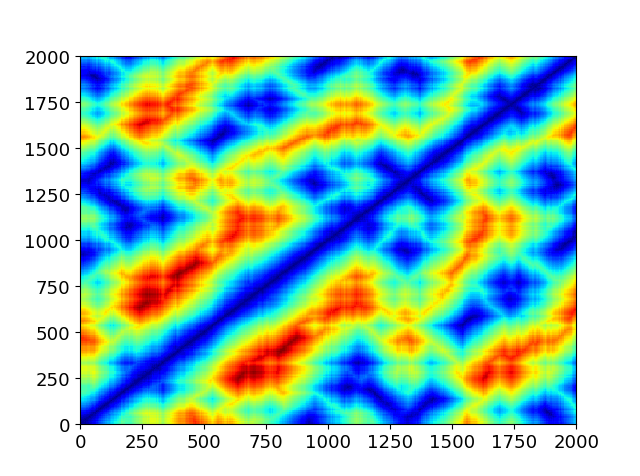}}
	\caption{Heatmaps for pairwise distance matrices.}
	\label{figheat}
	\vspace{-0.3cm}
\end{figure}
Note that we have removed the superscript $(i)$ and subscript $m$ for each term of $p_n$ and $t_n$ for brevity. Unlike the CIR geodesic distance proposed in \cite{frauhofer}, the proposed signature distance can be a faithful metric for both local and global dissimilarity without constructing any neighborhood graph.  Fig.\ref{figheat} shows some results about the heatmaps for the pairwise distance matrices of true radio environment, the proposed signature distance, and CIR geodesic distance with different neighborhood numbers, using the first 2K samples from the ultra wideband (UWB) dataset in \cite{frauhofer}. As can be seen, the heatmap of the pairwise signature distance shows similar pattern with that of physical distance. Whereas the result of the CIR geodesic distance can be seriously degraded when the neighborhood number $K$ is set inappropriately. With the distance metric for signature maps, the following loss function is proposed to train the Siamese network, 
\addtolength{\topmargin}{0.011in}
%\addtolength{\topmargin}{-0.005in}
%\addtolength{\topmargin}{0.006in}
\begin{equation}
\small{
	\mathcal{L}(\boldsymbol{o}_i,\boldsymbol{o}_j)=\sqrt{\sum_{(i,j)\in\mathcal{P}_{batch}}(\rho(\boldsymbol{\Lambda}_i,\boldsymbol{\Lambda}_j)-\rVert\boldsymbol{o}_i-\boldsymbol{o}_j\rVert_2)^2}
}
\vspace{-0.03in}
\end{equation}
where $\mathcal{P}_{batch}:=\{(i,j)|(i,j)\in\mathcal{D}\times\mathcal{D}\}$ is a set of index pairs with the cardinality being the batch number $|\mathcal{P}_{batch}|=N_{batch}$ and $\boldsymbol{o}_i$ denotes the low-dimensional embedding for the $i$-th sample. Note that we intentionally remove a scaling factor in the loss function, which is usually proportional to the batch number. Instead, we normalize the pairwise distance matrix, which can speed up the training time and shows better results.
%
%To obtain a measure for global similarity, the CIR geodesic distance need to be calculate, in which the neighbohood number need to be determined. 

%Moreover, we can further select a vector (the third column) from the signature map to perform PCA, which consists of the third-level signature feature and given by %the selected vector is given by 
%
%in which the closed form expression for the unnormalized $\boldsymbol{\lambda}_3^{(i,m)}$ is given by
%\begin{equation}
%	\lambda_3^{(i,m)}=
%\end{equation}
\begin{table}[htbp]
	\centering	
	\vspace{-0.06in}
	\caption{Simulation assumptions for InF-DH}
	\label{tb_infdh}
	\begin{center}
		\begin{tabular}{|c|c|}
			\hline
			\textbf{Parameter} & \textbf{Value}\\% & \textbf{TRP number} & \textbf{Antenna configuration}\\
			\hline 
			\text{Scenario} & {InF-DH}\\
			\hline
			\text{Hall size} & {$120\times60$ m}\\
			\hline
			\text{Carrier frequency} & {3.5GHz}\\
			\hline
			\text{Bandwidth} & 100MHz\\% & 0.06 & 0.09\\
			\hline
			\text{UE height}& 1.5m\\
			\hline
			\text{Clutter parameters} & density: 60\%, height: 6m, size: 2m\\
			\hline
			\text{Antenna configurations} & 1 element at BS and 1 element at UE,\\
			%			& isotropic antenna gain pattern\\
			\hline
			\text{Dimensionality of CIRs} & $18\times256\times2$\\
			\hline
			\text{Size of the dataset } & 6K for training and 2K for testing\\
			\hline 
		\end{tabular}
	\end{center}
\vspace{-0.5cm}
\end{table}
\begin{table}[htbp]
	\caption{Network architectures for the proposed SSNs}
	%\resizebox{1\textwidth}{1\textwidth}
	\vspace{-0.2cm}
	\begin{center}
		\begin{tabular}{|c|c|c|c|}
			\hline
			\textbf{Layer Type}&\multicolumn{2}{|c|}{\textbf{Output dimension}}&\multicolumn{1}{|c|}{\textbf{Kernel size}} \\
			\cline{2-4} 
			& \textbf{InF-DH}&\textbf{5G/UWB}&\textbf{InF-DH/5G/UWB}  \\
			\hline
			Conv& $16\times18\times6$ &$16\times6\times6$ &$3\times3$  \\
			\hline
			Conv&$32\times18\times6$&$32\times6\times6$&$5\times5$ \\
			\hline
			Pooling&$1\times18\times6$&$1\times6\times6$&-\\
			\hline
			Dense&10&10&-\\
			\hline
			Dense&2&2&-\\
			\hline
			%\multicolumn{6}{l}{$^{\mathrm{a}}$Sample of a Table footnote.}
		\end{tabular}
		\label{ssn_arc}
	\end{center}
\vspace{-0.5cm}
\end{table}
\begin{table}[ht]
	\caption{FLOPs comparison for different model input}
	%\resizebox{1\textwidth}{1\textwidth}
	\vspace{-0.2cm}
	\begin{center}
		\begin{tabular}{|c|c|c|c|}
			\hline
			\textbf{Input Type}&\multicolumn{3}{|c|}{\textbf{Dataset}} \\
			\cline{2-4} 
			& \textbf{InF-DH}&\textbf{5G}&\textbf{UWB}  \\
			\hline
			CIR magnitude & 60.53M &3.86M &15.76M  \\
			\hline
			Signature map (Ours)&1.42M&0.47M&0.47M \\
			\hline
			%\multicolumn{6}{l}{$^{\mathrm{a}}$Sample of a Table footnote.}
		\end{tabular}
		\label{tab_flops}
	\end{center}
	\vspace{-0.6cm}
\end{table}

\begin{table*}[ht]
	\vspace{0.06in}
	\caption{Experimental results for InF-DH dataset}
%	\resizebox{1\textwidth}{1\textwidth}
	\begin{center}
		\begin{tabular}{|c|c|c|c|c|c|c|c|c|}
			\hline
			\textbf{Method}&\multicolumn{4}{|c|}{\textbf{Training dataset}}&\multicolumn{4}{|c|}{\textbf{Testing dataset}}  \\
			\cline{2-9} 
			& \textbf{CT} $\uparrow$&\textbf{TW} $\uparrow$&\textbf{MAE}$\downarrow$ & \textbf{CE90} $\downarrow$& \textbf{CT} $\uparrow$& \textbf{TW} $\uparrow$& \textbf{MAE} $\downarrow$ & \textbf{CE90} $\downarrow$\\
			\hline
			\textbf{SPCA} (Ours)& 0.918 & 0.892 &  $14.73\pm0.16$ (m) & $25.97\pm0.32$ (m) & 0.914 & 0.892 &  $14.67\pm0.16$ (m)& $25.58\pm0.33$ (m) \\
			\hline
			CIR PCA& 0.949&0.961&$17.04\pm0.6$ (m)&$31.95\pm0.5$ (m)& 0.946&0.960&$17.04\pm0.6$ (m) &$31.95\pm0.5$ (m)\\
			\hline
			\textbf{FSSN} (Ours)&0.935&0.921&$13.78\pm0.16$ (m)&$24.55\pm0.29$ (m)&0.931&0.920&$13.54\pm0.15$ (m)&$23.88\pm0.33$ (m)\\
			\hline
			\textbf{PSSN} (Ours) &0.960&0.957&$12.43\pm0.16$ (m)&$23.11\pm0.33$ (m)&0.957&0.956&$12.44\pm0.15$ (m)&$22.67\pm0.34$ (m)\\
			\hline
			CIR Sia. & 0.941&0.935&$18.92\pm0.22$ (m)&$33.3\pm0.57$ (m)&0.941&0.938&$19.08\pm0.22$ (m)&$33.38\pm0.65$ (m)\\
			\hline
			%\multicolumn{6}{l}{$^{\mathrm{a}}$Sample of a Table footnote.}
		\end{tabular}
		\label{tinfdh}
	\end{center}
	\vspace{-0.2cm}
\end{table*}
\begin{table*}[h]
	\caption{Experimental results for 5G dataset}
	%\resizebox{1\textwidth}{1\textwidth}
	\begin{center}
		\begin{tabular}{|c|c|c|c|c|c|c|c|c|}
			\hline
			\textbf{Method}&\multicolumn{4}{|c|}{\textbf{Training dataset}}&\multicolumn{4}{|c|}{\textbf{Testing dataset}}  \\
			\cline{2-9} 
			& \textbf{CT} $\uparrow$&\textbf{TW} $\uparrow$&\textbf{MAE} $\downarrow$ & \textbf{CE90} $\downarrow$& \textbf{CT} $\uparrow$& \textbf{TW} $\uparrow$& \textbf{MAE} $\downarrow$ & \textbf{CE90} $\downarrow$\\
			\hline
			\textbf{SPCA} (Ours)& 0.906 & 0.839 & $2.00\pm0.10$ (m)& $3.63\pm0.21$ (m)& 0.908 & 0.845 & $1.93\pm0.09$ (m)& $3.54\pm0.21$(m) \\
			\hline
			CIR PCA& 0.911&0.855&$4.25\pm0.07$ (m) &$7.66\pm0.17$ (m)& 0.914&0.864 &$4.11\pm0.05$ (m)&$7.37\pm0.13$ (m)\\
			\hline
			\textbf{FSSN} (Ours)&0.915&0.843&$4.59\pm0.04$ (m)&$7.74\pm0.17$ (m)&0.919&0.855&$4.27\pm0.04$ (m)&$7.61\pm0.18$ (m)\\
			\hline
			\textbf{PSSN} (Ours)&0.987&0.986&$1.30\pm0.02$ (m)&$2.24\pm0.03$ (m)&0.984&0.983&$1.34\pm0.02$ (m)&$2.40\pm0.03$ (m)\\
			\hline
			CIR Sia.& 0.980&0.979&$1.54\pm0.02$ (m)&$2.74\pm0.04$ (m)&0.976&0.974&$1.60\pm0.02$ (m)&$2.93\pm0.04$ (m) \\
			\hline
			CIR Sia. in \cite{frauhofer}&0.986&0.986&1.40 (m)&2.35 (m)&0.983&0.982&1.46 (m)&2.48 (m)\\
			\hline
			%\multicolumn{6}{l}{$^{\mathrm{a}}$Sample of a Table footnote.}
		\end{tabular}
		\label{t5g}
		\vspace{-0.2cm}
	\end{center}
\end{table*}
\begin{table*}[h]
	\caption{Experimental results for UWB dataset}
	%\resizebox{1\textwidth}{1\textwidth}
	\vspace{-0.2cm}
	\begin{center}
		\begin{tabular}{|c|c|c|c|c|c|c|c|c|}
			\hline
			\textbf{Method}&\multicolumn{4}{|c|}{\textbf{Training dataset}}&\multicolumn{4}{|c|}{\textbf{Testing dataset}}  \\
			\cline{2-9} 
			& \textbf{CT} $\uparrow$&\textbf{TW} $\uparrow$&\textbf{MAE} $\downarrow$ & \textbf{CE90} $\downarrow$& \textbf{CT} $\uparrow$& \textbf{TW} $\uparrow$& \textbf{MAE} $\downarrow$ & \textbf{CE90} $\downarrow$\\
			\hline
			\textbf{SPCA} (Ours)& 0.976 & 0.962 & $1.52\pm0.02$ (m)&$2.75\pm0.04$ (m)& 0.976 & 0.965 &  $1.46\pm0.02$ (m)& $2.72\pm0.04$ (m)\\
			\hline
			CIR PCA& 0.934&0.825&$3.79\pm0.07$ (m)&$7.49\pm0.22$ (m)& 0.941&0.874&$3.51\pm0.05$  (m)&$6.33\pm0.14$ (m)\\
			\hline
			\textbf{FSSN} (Ours) &0.980&0.970&$1.33\pm0.01$ (m)&$2.35\pm0.02$ (m)&0.978&0.970&$1.33\pm0.01$ (m)&$2.38\pm0.03$ (m)\\
			\hline
			\textbf{PSSN} (Ours)&0.997&0.996&$0.66\pm0.01$ (m)&$1.24\pm0.02$ (m)&0.996&0.996&$0.67\pm0.01$ (m)&$1.24\pm0.02$ (m)\\
			\hline
			CIR Sia.& 0.992&0.991&$0.88\pm0.01$ (m)&$1.60\pm0.03$ (m)&0.991&0.990&$0.87\pm0.01$ (m)&$1.58\pm0.02$ (m)\\
			\hline
			CIR Sia. in \cite{frauhofer}&0.997&0.997&0.69 (m)&1.30 (m)&0.997&0.996&0.72 (m)&1.28 (m)\\
			\hline
			%\multicolumn{6}{l}{$^{\mathrm{a}}$Sample of a Table footnote.}
		\end{tabular}
		\label{tab_uwb}
	\end{center}
\vspace{-0.5cm}
\end{table*}
\section{Experimental results}\label{sec4}
The proposed method is evaluated in one synthetic dataset in the scenario of  indoor factory with dense clutter and high base station (InF-DH) from the 3rd Generation Partnership Project (3GPP)\cite{38901} and two open-source real-filed datasets for 5G and ultra-wideband (UWB) systems \cite{dataset}. The simulation assumptions for the InF-DH dataset are given in Table \ref{tb_infdh}. The configurations of 5G and UWB datasets can be found in \cite{dataset}. We use the preprocessed samples for UWB and 5G datasets \cite{frauhofer}, in which  the time of arrival (TOA) or the time difference of arrival (TDOA)  information is embedded into the CIR. For all datasets, every sample is transformed to a $N_b\times L$ signature map with $N_b=18$ for InF-DH, $N_b=6$ for 5G and UWB datasets, and $L=6$ representing the number of signature features from the 2nd to the 4th level. The first level signature features (which are $c_N$ and $t_N$) are discarded since they are uninformative after normalizing the CIR energy. Compared with the dimensionality of CIRs, the feature number has been reduced more than 87\% for 5G and 97\% for InF-DH and UWB datasets,  which means a significant reduction in complexity for subsequent procedures of CC.

 %Each sample consists of CIRs from 18 BSs with the dimensionality of $18\times256\times2$ (256 denotes the number of CIR taps, and 2 indicates real and imaginary parts).  The 5G dataset consists of about 18K samples for training and 15K for testing, and each CIR length is 49. The UWB dataset consists of about 18K samples for training and 3K for testing, and each CIR length is 200. 

%For all three datasets, we transform each CIR into 6 signature features from the 2nd to the 4th level. The resulting  2D signature map with dimension of $N_b\times 6$ is fed into the SSN. 
%\vspace{-0.3in}
The performance of local similarity for CC is evaluated in terms of the continuity (CT) and trustworthiness (TW)  \cite{cc}. The value of these two metrics ranges from 0 to 1, and neighbor  points that are embedded far away from each other decrease CT, whereas far away points that are embedded as neighbor points decrease TW. For measuring global dissimilarity, we perform an affine transformation from the chart to real coordinates and evaluate the mean absolute errors (MAE) and the 90th-percentile of the cumulative distribution function of the error (CE90) in units of meters.  We randomly choose 100 labeled samples to obtain the least-squares estimation of the affine  matrix and perform 1K experiments to obtain the mean and standard deviation of MAE and CE90. 
Moreover, the proposed SPCA and SSN are compared with CIR based methods including CIR magnitude based PCA or Siamese network \cite{frauhofer}. For SSN based CC, we further evaluate two approaches, namely the full SSN (FSSN)  and partial SSN (PSSN) based charting. FSSN uses the signature map as network input and also uses it to produce the pairwise distance matrix by \eqref{eq:dist_def}, and the PSSN only uses the signature map as network input and the pairwise distance matrix is generated by calculating CIR geodesic distance. Moreover, the Adam optimizer is used for training all networks with 50 epochs, and the batch size is set to be 50 for InF-DH and 500 for both 5G and UWB datasets, and the learning rate is set to be 1e-4 for InF-DH and 1e-3 for both 5G and UWB datasets.% accordingly.

%\vspace{-1in}
The network architectures and output dimensions for each layer are given in Table \ref{ssn_arc}. The model consists of two Conv layers, one adaptive average pooling, and two dense layers. Each Conv layer consists of a 2-D convolutional layer, a batch normalization (BN) and a  rectified linear unit (ReLU) as the activation function. The number of floating-point operations (FLOPs) for the network with CIRs or the signature map as model input are given in Table \ref{tab_flops}. It can be seen that the number of FLOPs is reduced significantly for all datasets. 
%
%Despite these datasets have different CIR lengths, we transform all the original data sample into a $N_b\times 6$ signature map with $N_b=18$ for InF-DH dataset and $N_b=6$ for 5G and UWB datasets.
%\addtolength{\topmargin}{-0.01in}
\subsection{Results for InF-DH dataset}
\begin{figure}[h]
	\vspace{-0.37cm}
%	\vspace{0.03in}
	\centering
	\subfloat[True UE locations]{
		\includegraphics[width=2.7cm]{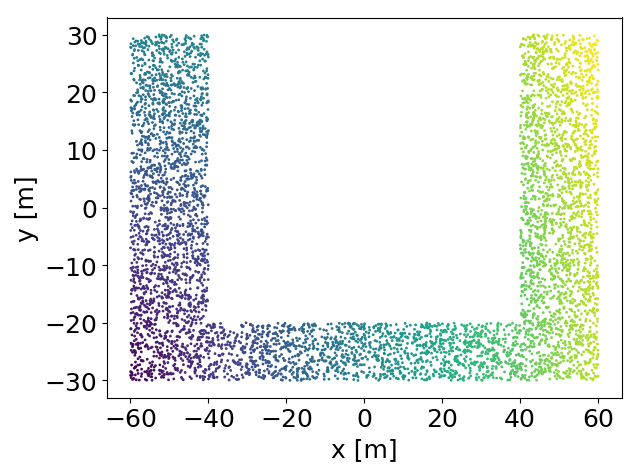}}
	\subfloat[SPCA]{
		\includegraphics[width=2.7cm]{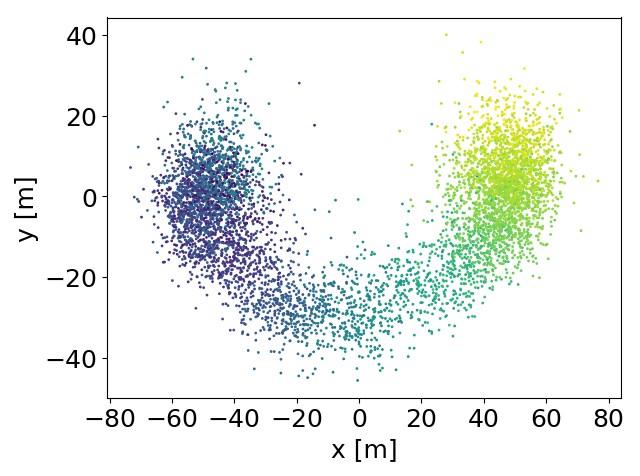}}
	\subfloat[CIR based PCA]{
		\includegraphics[width=2.7cm]{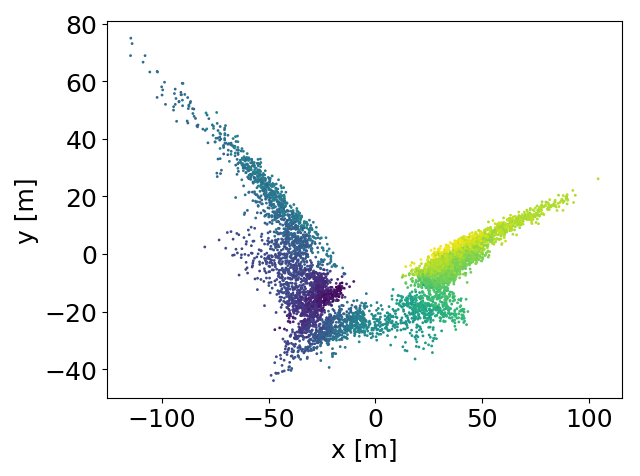}}
	
	\vspace{-0.4cm}
	\subfloat[FSSN]{
		\includegraphics[width=2.7cm]{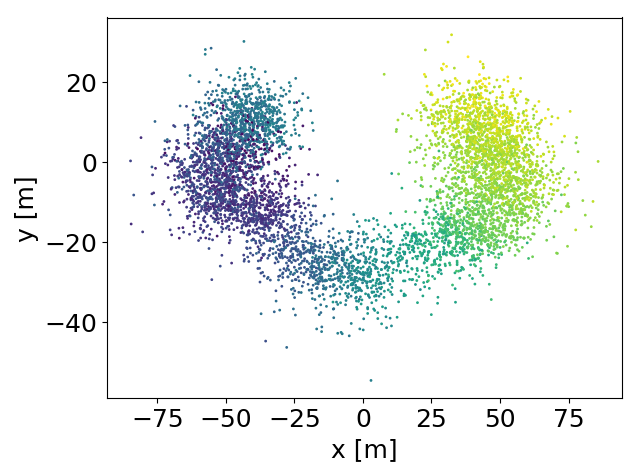}}
	\subfloat[PSSN]{
		\includegraphics[width=2.7cm]{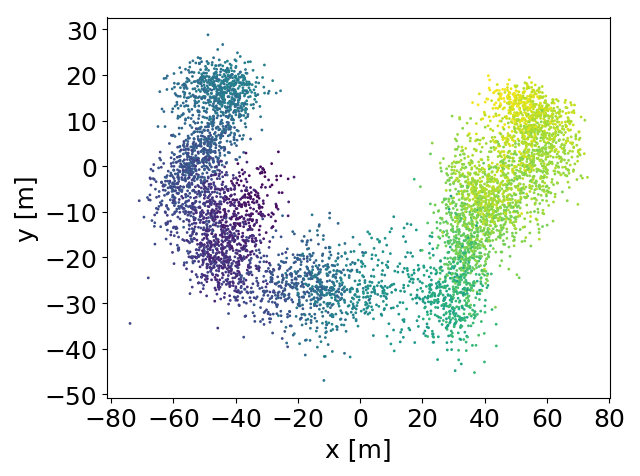}}
	\subfloat[CIR based Sia.]{
		\includegraphics[width=2.7cm]{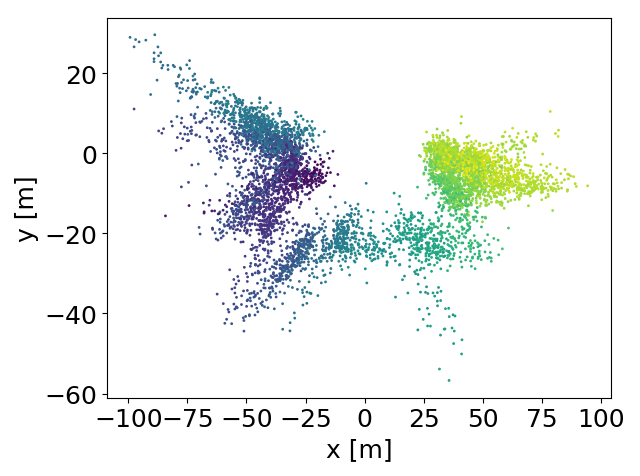}}
	\vspace{-0.1cm}
	\caption{Ground-truth UE locations and CC results after affine transformation for InF-DH dataset.}
	\label{ccinfdh}
	\vspace{-0.3cm}
\end{figure}

%For SPCA, as descriped in Section.\ref{sec3b},
The true UE locations and the charting results after affine transformation are given in Fig.\ref{ccinfdh}, and detailed results for all evaluated methods are given in Table \ref{tinfdh}. For PCA based CC, we select the vector $\boldsymbol{s}_i$ from the signature map for each sample to do PCA, which means we only need to do SVD for an $18\times18$ covariance matrix rather than a $4608\times4608$ matrix for CIR based PCA. Moreover, SPCA shows better results on global performance in terms of MAE and CE90 and slightly degraded local similarity in terms of CT and TW, compared with CIR based PCA. For Siamese network based CC, both FSSN and PSSN show better results on global performance and similar results on local similarity in comparison of CIR based method. It thus implies that better performance and significant complexity reduction can be achieved simultaneously with our proposed method. 
%
%The PSSN shows better results on both local and global similarity in comparison of FSSN. The reason is that using a carefully chose number of neighborhood to constructing the CIR geodesic distance matrix can indeed improve the distance matrix design but at the cost of more computational complexity.
%For PSSN and CIR based Siamese network, the neighborhood number for the CIR geodesic distance is set to be 20.
 \begin{figure}[h]
	\vspace{-0.4cm}
	\centering
	\subfloat[True UE locations]{
		\includegraphics[width=2.7cm]{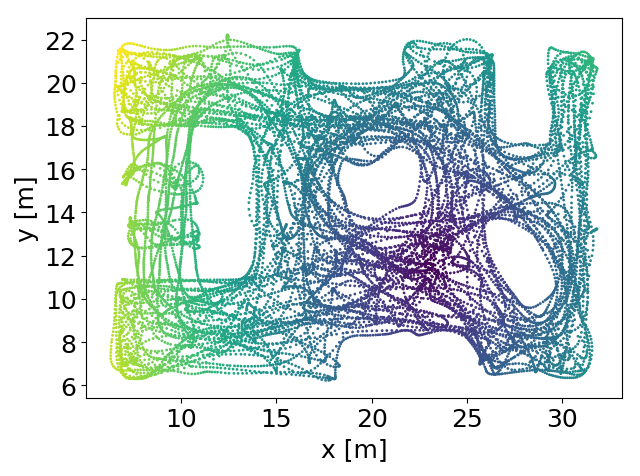}}
	\subfloat[SPCA]{
		\includegraphics[width=2.7cm]{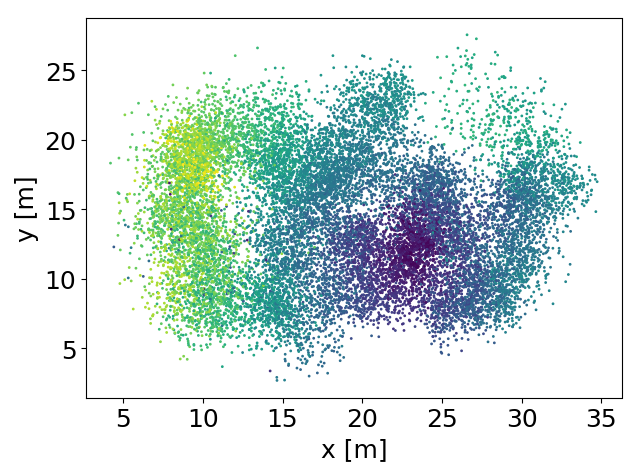}}
	\subfloat[CIR based PCA]{
		\includegraphics[width=2.7cm]{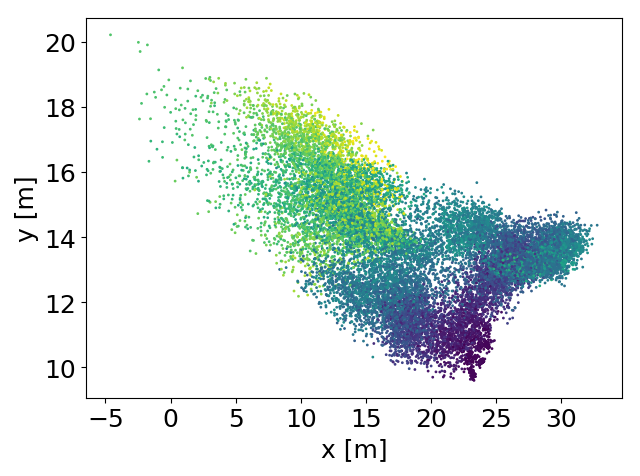}}
	
	\vspace{-0.4cm}
	\subfloat[FSSN]{
		\includegraphics[width=2.7cm]{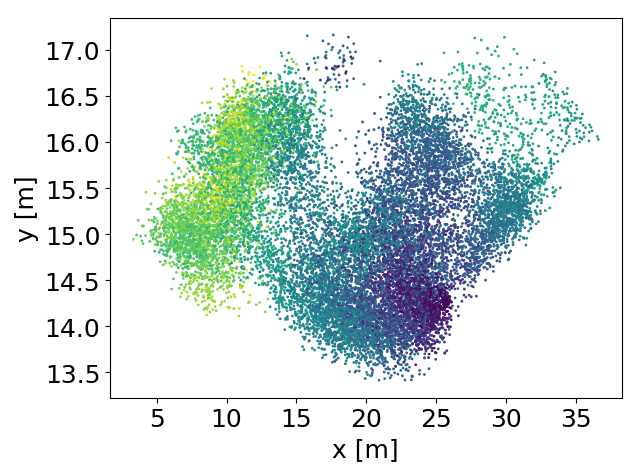}}
	\subfloat[PSSN]{
		\includegraphics[width=2.7cm]{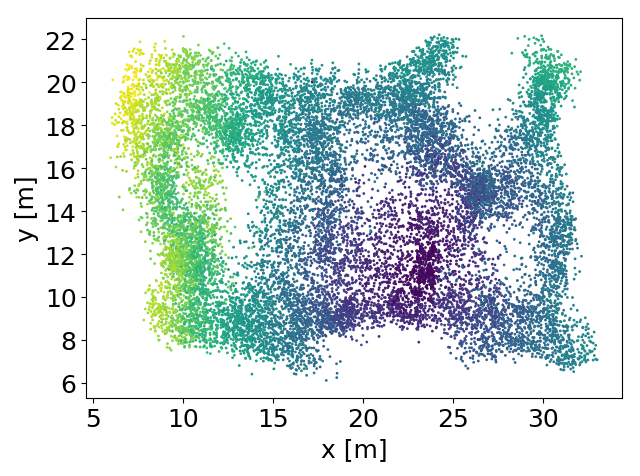}}
	\subfloat[CIR based Sia.]{
		\includegraphics[width=2.7cm]{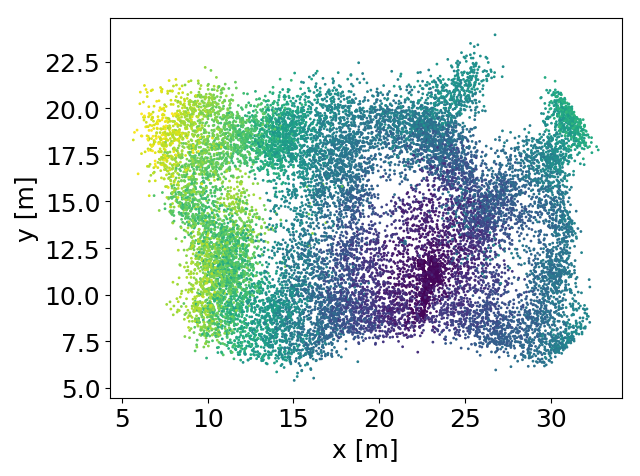}}
		\vspace{-0.1cm}
	\caption{Ground-truth UE locations and CC results after affine transformation for 5G dataset.}
	\label{cc5g}
	\vspace{-0.7cm}
\end{figure}
\begin{figure}[h]
	\vspace{-0.37cm}
%	\vspace{0.03in}
	\centering
	\subfloat[True UE locations]{
		\includegraphics[width=2.7cm]{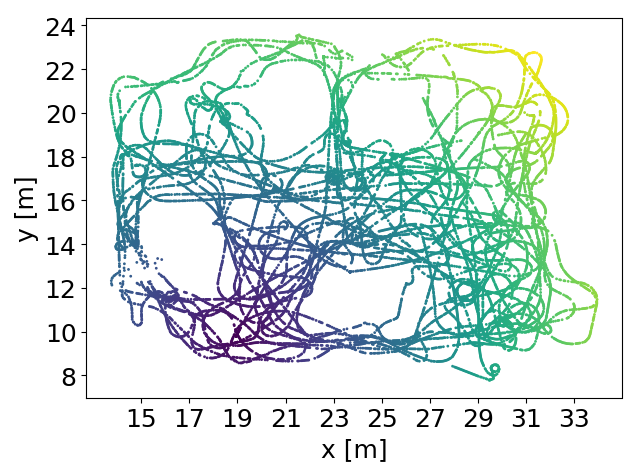}}
	\subfloat[SPCA]{
		\includegraphics[width=2.7cm]{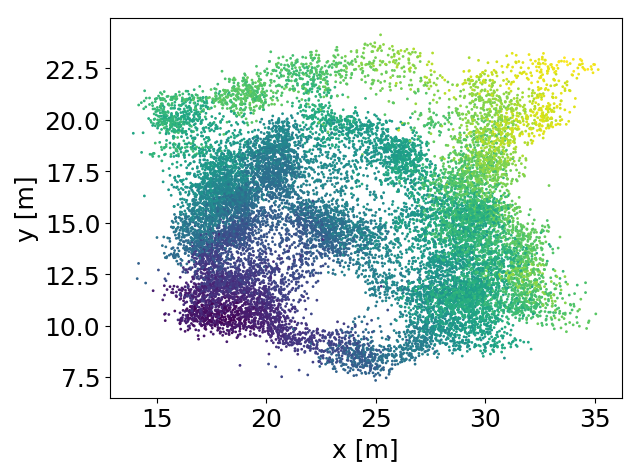}}
	\subfloat[CIR based PCA]{
		\includegraphics[width=2.7cm]{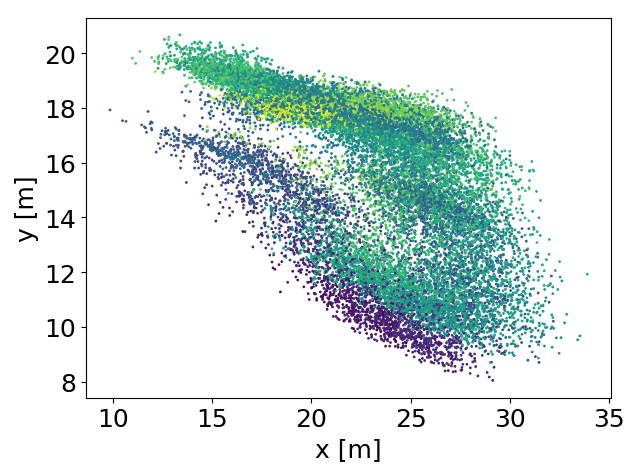}}
	
	\vspace{-0.4cm}
	\subfloat[FSSN]{
		\includegraphics[width=2.7cm]{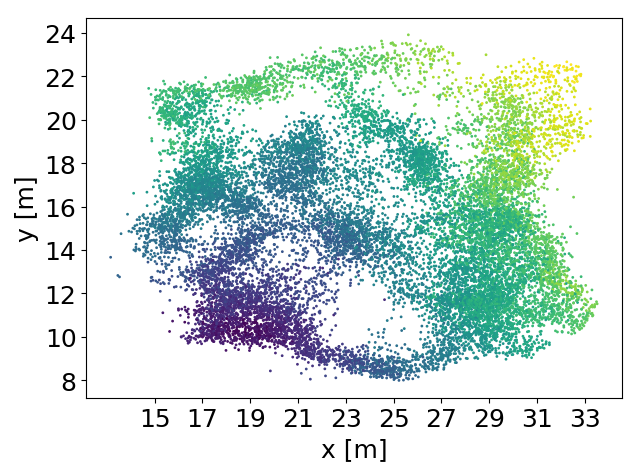}}
	\subfloat[PSSN]{
		\includegraphics[width=2.7cm]{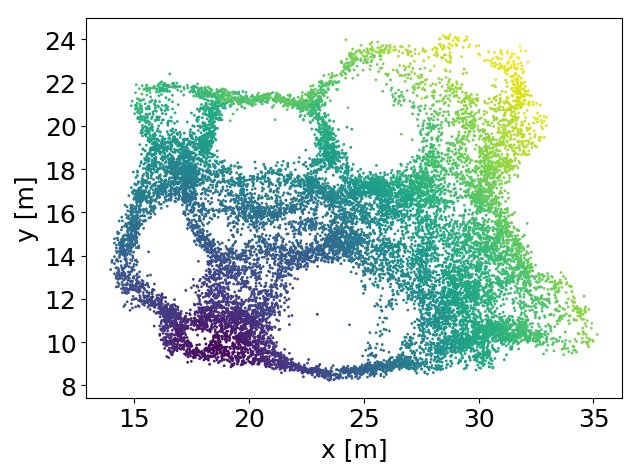}}
	\subfloat[CIR based Sia.]{
		\includegraphics[width=2.7cm]{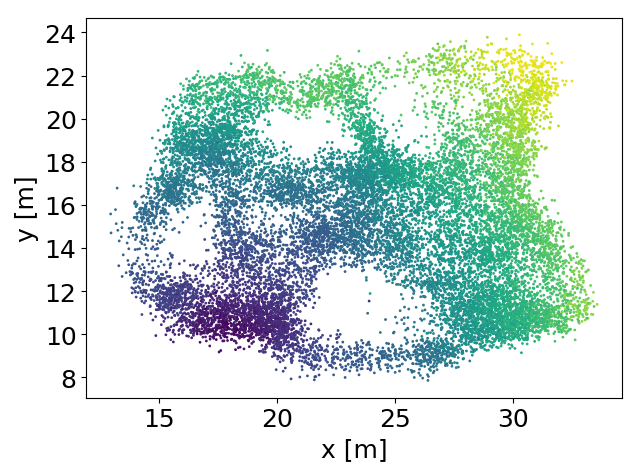}}
		\vspace{-0.1cm}
	\caption{Ground-truth UE locations and CC results after affine transformation for UWB dataset.}
	\label{cc_uwb}
	\vspace{-0.5cm}
\end{figure}
\addtolength{\topmargin}{0.037in}
\subsection{Results for 5G and UWB datasets}
For the 5G dataset, the experimental results are given in Table \ref{t5g} and Fig.\ref{cc5g}. For PCA based method, the vectorized signature map is used to obtain the covariance matrix and more  components are used for the affine transformation. It can be seen that SPCA shows much better results on global similarity in terms of MAE and CE90 compared with CIR based PCA. For Siamese network based charting, the result reported in \cite{frauhofer} is also given, which is obtained by a more complex Siamese network and using all training samples to perform the affine transformation. It shows that PSSN achieves the best results for both local and global similarity among all methods, with a complexity reduction about 88\% (as shown in Table \ref{tab_flops}). Moreover, the results of FSSN are unexpectedly worse than PSSN and CIR based methods. The reason is that only TDOA information is embeded in the CIR and the reference BS for the measurement of TDOA is not even fixed, which decreases the reliability for the proposed signature distance metric.  %some local and global similarity when using the proposed distance metric for the signature map. %Because there is only TDOA information in each CIR for this dataset and the reference BS depends on perticular UE locations. 

For the UWB dataset, experimental results are given in Table\ref{tab_uwb} and Fig.\ref{cc_uwb}. For PCA based method, due to the TOA information is kept in each CIR, we select the vector
$\boldsymbol{s}_i$ for constructing the covariance matrix just like the SPCA for Inf-DH dataset,  which can further reduce the complexity without sacrificing the performance. It can be seen that both the local and global performance are greatly improved when using SPCA instead of CIR based PCA, and the performance can be further improved by FSSN. Compared with the result from  \cite{frauhofer}, PSSN achieves better global similarity and almost the same local similarity, with a complexity reduction more than 97\%.

\section{Conclusions}
In this paper, a novel signature based approach for global CC is proposed, in which a compact signature map is extracted from CIRs and can be used for PCA or Siamese network based CC. Moreover, a novel distance metric, which dose not need to construct neighborhood graph, is proposed for training Siamese networks. The experimental results show that the proposed approach can achieve better  performance on global and local similarity than CIR based methods, with significantly reduced complexity.

%novel dimensionality reduction technique is proposed for
%AI-based positioning, which uses log signature-based operation
%to extract features of wireless channels. With its elegant
%feature extraction capability, about 98 percent reduction can be
%achieved, regarding the input feature number and the computational
%complexity of the subsequent AI/ML model used for
%positioning, without compromising the positioning accuracy.
%Meanwhile, better positioning accuracy can be achieved even
%with simple AI/ML models and limited training samples

%\vspace{12pt}
%\color{red}
%IEEE conference templates contain guidance text for composing and formatting conference papers. Please ensure that all template text is removed from your conference paper prior to submission to the conference. Failure to remove the template text from your paper may result in your paper not being published.
\end{document}